\documentclass[nopubldata]{lpor2014}
\usepackage{blindtext}
\usepackage{hyperref}
\hypersetup{%
  colorlinks,
  plainpages=false,
  pdfpagelabels,
  breaklinks,
  pdfview=Fit,
  bookmarksopen,
  bookmarksnumbered,
  linkcolor=black,
  anchorcolor=black,
  citecolor=black,
  filecolor=black,
  urlcolor=LPRred
  }%
\graphicspath{{./figs/}}
\setpages[]{}
\setvolume[0]{0}
\setyear{2011}%
\setdoi{201100000}%
\oddheadlogotext{}{Guide}
\begin{document}

\titlefigure{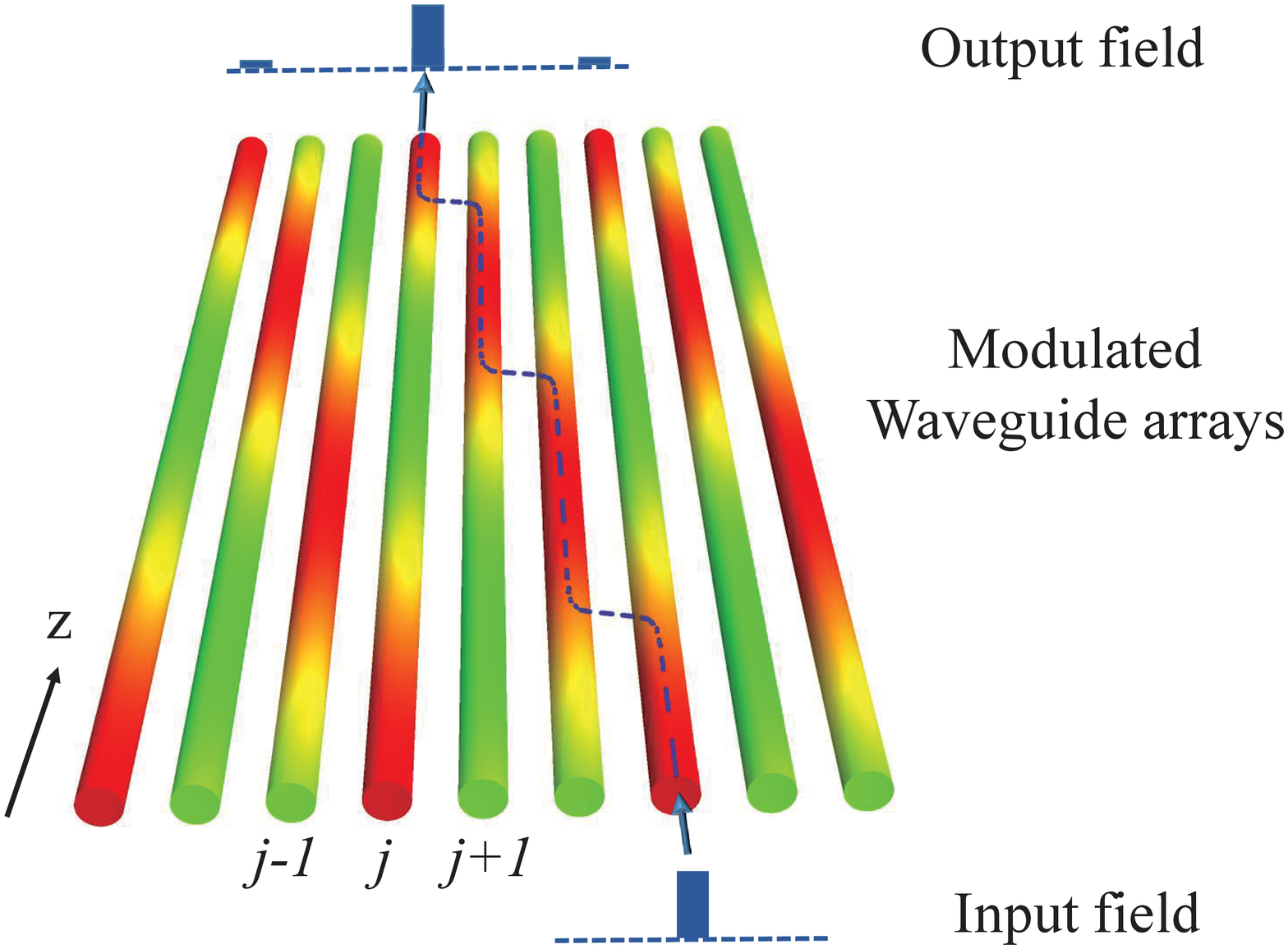}
\abstract{
Photonic waveguide arrays provide an excellent platform for simulating conventional topological systems, and they can also be employed for the study of novel topological phases in photonics systems. However, a direct measurement of bulk topological invariants remains a great challenge. Here we study topological features of generalized commensurate Aubry-Andr\'{e}-Harper (AAH) photonic waveguide arrays and construct a topological phase diagram by calculating all bulk Chern numbers, and then explore the bulk-edge correspondence by analyzing the topological edge states and their winding numbers. In contrast to incommensurate AAH models, diagonal and off-diagonal commensurate AAH models are not topologically equivalent. In particular, there appear nontrivial topological phases with large Chern numbers and topological phase transitions. By implementing Thouless pumping of light in photonic waveguide arrays, we propose a simple scheme to measure the bulk Chern numbers.
}

\title{Topological Phase Transitions and Thouless Pumping of Light in Photonic Waveguide Arrays}

\author{Yongguan Ke\inst{1,2}, Xizhou Qin\inst{1}, Feng Mei\inst{1}, Honghua Zhong\inst{1}, Yuri S. Kivshar\inst{3}, and Chaohong Lee\inst{1,2,3,*}}%
\authorrunning{Y. Ke et al.}
\mail{lichaoh2@mail.sysu.edu.cn; chleecn@gmail.com}

\institute{%
TianQin Research Center \& School of Physics and Astronomy, Sun Yat-Sen University (Zhuhai Campus), Zhuhai 519082, China
\and
State Key Laboratory of Optoelectronic Materials and Technologies, Sun Yat-Sen University (Guangzhou Campus), Guangzhou 510275, China
\and
Nonlinear Physics Centre, Research School of Physics and Engineering,
Australian National University, Canberra ACT 2601, Australia}


\maketitle

\section{Introduction}
\label{sec:intro}

Photonic systems represent a promising platform for testing quantum principles and implementing quantum simulations~\cite{Longhi2009,Aspuru2012}.
The main advantages of photonic systems stem from: (i) easily preparing and detecting states, (ii) manipulating dispersion of light by photonic structures,
(iii) directly visualizing dynamics in space without decoherence, and (iv) exploring dynamics regime exceeding original systems. Photonic toolbox has been used to mimic various exotic quantum phenomena, such as,  quantum chemistry~\cite{Lanyon2010,Ma2011}, Bloch-Zener dynamics~\cite{Morandotti1999,Pertsch1999,Dreisow2009}, Anderson localization~\cite{Schwartz2007,Segev2013}, and dynamics localization~\cite{Longhi2006,Szameit2009a}.
Recently, photonic systems have been used to simulate conventional topological systems and explore new topological phases~\cite{Lu2014, Li2009, Rechtsman2013, Lumer2013, Zeuner2015, Karzig2015, Bliokh2015, Slobozhanyuk2015}. The integer quantum Hall systems have been simulated by magneto-optical photonic crystals and coupled ring resonators~\cite{Klitzing1980,Wang2009,Hafezi2011}. It has been found that topological edge states of light will propagate along the boundary~\cite{Wang2009,HafeziM2013} in the presence of internal and external disorder.

Modulating photonic waveguide arrays~\cite{Kraus2012a}, one may mimic Aubry-Andr\'{e}-Harper (AAH) models~\cite{Harper1955,Aubry1980,Hatsugai1990,Kraus2012a,Kraus2012b,HLu2015,Liu2015}, which are equivalent to two-dimensional integer quantum Hall systems~\cite{Harper1955,Aubry1980}.
In particular, strong hopping modulation which is unlikely to be accessible in electronic systems, can be realized in photonic systems.
For incommensurate systems, regardless of whether the quasiperiodic modulation is applied to on-site potential or off-site hopping, their topological equivalence has been found in theory~\cite{Kraus2012b}.
For commensurate systems, topological phase transitions have been explored for some specific parametric conditions ~\cite{Hatsugai1990}, and the topological zero-energy modes have been discovered in the gapless regime~\cite{Ganeshan2013}.
However, the existence of topological phases in the whole parameter space are still unclear, in particular the phase diagram has never been given.

By employing the bulk-edge correspondence, one may explore topological phases by probing edge states or edge topological invariants~\cite{Meidan2011,Fulga2012,Hu2015,Mittal2016,Hafezi2014,Poshakinskiy2015}.
In recent years, in addition to observing edge states and adiabatic pumping~\cite{Kraus2012a, HLu2015}, the edge topological invariant has also been measured~\cite{Hu2015,Mittal2016}. However, direct measurement of the bulk topological invariants such as Chern number is still a great challenge in photonic systems~\cite{Ozawa2014,Bardyn2014}. Due to the absence of any exclusion principle for photons, measuring bulk topological invariants need a complex procedure of sweeping the wavefunction over through the Brillouin zone~\cite{Atala2013,Aidelsburger}. That is, one has to know detailed information about initial states of the occupied bulk band. It has been proposed that the Thouless pumping is an alternative method for measuring bulk topological invariants~\cite{Thouless1983}. Recently, the topological Thouless pumping of ultracold atoms in optical superlattices has been reported in experiments~\cite{Nakajima2016,Lohse2015}.
Thus, we wonder \emph{if the Thouless pumping of light can be employed to explore nontrivial topological phases in photonic waveguide arrays}?

Here we study topological features of generalized commensurate AAH photonic waveguide arrays. By varying the hopping amplitude and on-site potential, we discover the topological phase diagram characterized by different Chern numbers. Interestingly, nontrivial topological phases with large Chern numbers and topological phase transitions appear when the hopping modulation is sufficiently strong. This means that diagonal and off-diagonal commensurate AAH models may have different topological invariants. To measure Chern numbers, in addition to the transverse modulations along the lattice direction, we propose to apply periodic longitudinal modulations along the propagation direction for inducing Thouless pumping of light along the transverse direction. Therefore, in the adiabatic limit, the Chern number of the occupied band can be determined by measuring the mean transverse position shift of light in one longitudinal modulation period.

\section{Model}
\begin{figure}[!htp]
\begin{center}
\includegraphics[width=1\columnwidth]{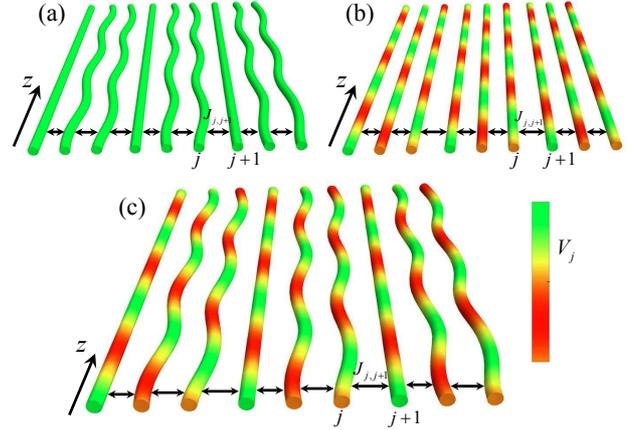}
\end{center}
\caption{ Schematic of different types of photonic waveguide arrays: (a) off-diagonal AAH model, (b) diagonal AAH model, and (c) the generalized AAH model. The on-site potentials $V_j$ (denoted by color) and the hopping strengths $J_{j,j+1}$ can be modulated by controlling the refractive index and the inter-waveguide separation, respectively. The modulation phase $k_y$ is varying along the longitudinal propagation direction $z$.}
\label{diagram}
\end{figure}

We consider a generalized 1D commensurate AAH model obeying a single-particle single-band and tight-binding Hamiltonian
\begin{equation}
\label{GAAH}
\hat H = \sum\limits_{j = 1}^N {\left(J_{j,j+1} c_j^\dag {c_{j + 1}} + h.c.\right)} +\sum \limits_{j = 1}^N {V_j c_j^\dag {c_j}}.
\end{equation}
with the on-site potential $V_j= {\nu _d}\cos (2\pi \beta j + {k_y} )$ and the cosine-modulated nearest-neighbouring hopping $J_{j,j+1} = - J + {\nu_{od} }\cos (2\pi \beta j + k_y+\delta \phi)$, where the rational parameter $\beta=p/q$ ($p$ and $q$ are coprime numbers).
Here, $N=qL$ is the total number of lattice sites, $j$ denotes the lattice index, ${c_j}$ and $c_j^\dag$ are respectively annihilation and creation operators for the $j$-th site.
The diagonal modulations $V_j = {\nu _d}\cos (2\pi \beta j + {k_y})$ are described by the amplitude $\nu_{d}$, the frequency $\beta$ and the phase $k_y$.
The off-diagonal modulations ${\nu_{od} }\cos (2\pi \beta j + k_y +\delta \phi)$ are described by the amplitude $\nu_{od}$, the frequency $\beta$ and the phase $k_y+\delta \phi$.
Here $\delta \phi$ is the relative phase between the diagonal and off-diagonal modulations.
In the cases of $\nu_{od}=0$ and $\nu_{d}=0$, the generalized AAH model is reduced to the diagonal and off-diagonal AAH models, respectively.
Our AAH-type model with the tunable parameter $k_y$ can be realized by 1D photonic waveguide arrays, see Fig.~\ref{diagram}.
Since the on-site potentials are determined by the refractive indices and the inter-waveguide coupling drops exponentially with the inter-waveguide separation,
the diagonal and off-diagonal modulations can be realized by controlling the refractive indices and the inter-waveguide separation, respectively.

Imposing the periodic boundary condition on the system, its Hamiltonian can be block diagonalized as $\hat H = \bigoplus \hat H_{k_x}$ with the good quantum number $k_x = 2\pi l/(qL)$ for integers $l=\{1, 2, \cdots, L\}$.
Similar to other 1D systems~\cite{Lang2012,Zhu2013}, the tunable parameter $k_y$ provides another synthetic dimension, thus the decoupled blocks of our Hamiltonian are given as
\begin{equation}
  \hat H_{{k_x},k_y} = \sum\limits_{j = 1}^q {\left(J_{j,j+1} {\textrm{e}^{i{k_x}}}c_{j}^\dag {c_{j + 1}} + h.c.\right)}+
  \sum\limits_{j=1}^q{V_j c_j^\dag {c_j}}.
\end{equation}
Thus, the system becomes an effective 2D system with $q$ bands and it may have nontrivial Chern numbers over the Brillouin-like zone $(-\pi/q <k_x \le \pi/q, 0<k_y\le 2\pi)$.

\section{Topological phase transitions}
\label{ssec:preamble}

The band structure can be obtained by solving
$\hat H_{{k_x},{k_y}}\left| {{\psi_{n}}} \right\rangle  = {E_{n,k_x,k_y}}\left| {{\psi _n}} \right\rangle$.
The Chern number for the $n$-th band is defined as
\begin{equation}
  {\mathcal{C}_n} = \frac{1}{{2\pi}}\int_{ - \pi /q}^{\pi /q} {\textrm{d}{k_x}\int_0^{2\pi } {\textrm{d}{k_y}\mathcal{F}_{n}(k_x,k_y)} },
\end{equation}
where ${\mathcal{F}_{n}} = \operatorname{Im} \left( {\left\langle {{{\partial _{{k_y}}}{\psi _n}}}
 \mathrel{\left | {\vphantom {{{\partial _{{k_y}}}{\psi _n}} {{\partial _{{k_x}}}{\psi _n}}}}
 \right. \kern-\nulldelimiterspace}
 {{{\partial _{{k_x}}}{\psi _n}}} \right\rangle  - \left\langle {{{\partial _{{k_x}}}{\psi _n}}}
 \mathrel{\left | {\vphantom {{{\partial _{{k_x}}}{\psi _n}} {{\partial _{{k_y}}}{\psi _n}}}}
 \right. \kern-\nulldelimiterspace}
{{{\partial _{{k_y}}}{\psi _n}}} \right\rangle } \right)$  is the Berry curvature for the eigenstate $\left| {{\psi _n}} \right\rangle$.
Below we only consider commensurate AAH models with odd $q$ (such as $\beta=p/q=1/3$) and $\delta \phi=0$.
For the off-diagonal AAH model of rational parameter $\beta=1/(2q)$, the zero-energy edge modes appear in the gapless regime~\cite{Ganeshan2013}, but their Chern numbers are not well defined.

\begin{figure}[!htp]
\begin{center}
\includegraphics[width=1\columnwidth]{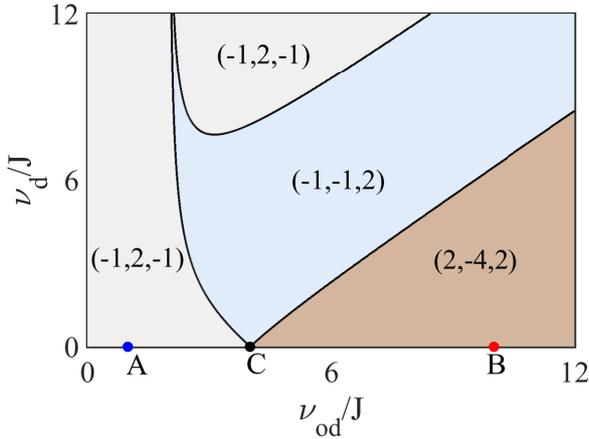}
\end{center}
\caption{ Topological phase diagram in the parameter plane of $(\nu_{od},\nu_d)$ for the generalized AAH model with $\beta=1/3$.}
\label{phase_transition}
\end{figure}

By using a manifestly gauge-invariant description~\cite{Takahiro2005}, we numerically calculate the Chern numbers of the three energy bands in the discretized Brillouin-like zone.
Given $\beta=1/3$, Fig.~\ref{phase_transition} shows the topological phase diagram in the parametric plane of $(\nu_{od}, \nu_{d})$.
The three numbers in the phase diagram are the Chern numbers of the three energy bands.
In the absence of hopping modulation ($\nu_{od}=0$), the system becomes a diagonal AAH model, which does not show topological phase transition for a given $\beta$~\cite{Thouless1982}.
For weak hopping modulations, similar to the incommensurate systems~\cite{Kraus2012b}, the topological phase remains the same Chern numbers because the band gaps remain open.

\begin{figure}[htp]
\begin{center}
\includegraphics[width=1\columnwidth]{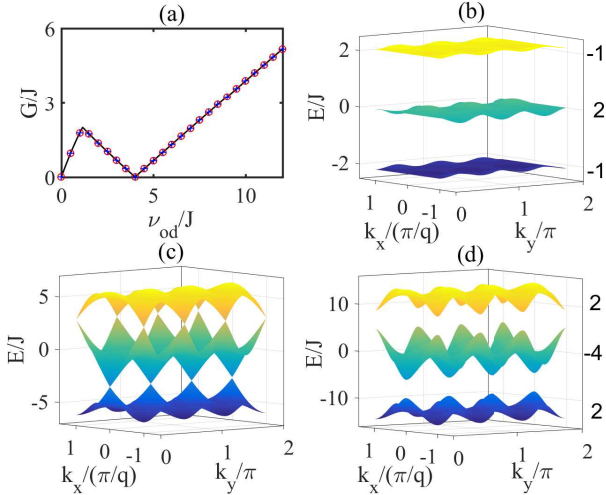}
\end{center}
\caption{ Energy gaps and bands for the off-diagonal AAH model with $\beta =1/3$ ($p=1$ and $q=3$).
(a) The two energy gaps as a function of the relative hopping modulation strength $\nu_{od}/J$. The blue `+' and red circle stands for the first and second gaps, respectively.
(b) Energy bands for $\nu_{od}/J=1$. The corresponding Chern numbers for the three bands are $(-1,2,-1)$.
(c) Energy bands for $\nu_{od}/J=4$. The two gaps close at the same point and there is no well definition for the Chern number.
(d) Energy bands for $\nu_{od}/J=10$. The corresponding Chern numbers for the three bands are $(2,-4,2)$.}
\label{gap}
\end{figure}

Different from the incommensurate systems~\cite{Kraus2012b}, topological phase transitions appear when the hopping modulation is sufficiently strong.
In the absence of diagonal modulation ($\nu_d=0$), the system is an off-diagonal AAH model with chiral symmetry~\cite{Kondakci2015}.
As topological phase transitions associate with band gap closure, we calculate the energy gaps as a function of the relative modulation amplitude $\nu_{od}/J$, see Fig.~\ref{gap}~(a).
The energy gap between the $n$-th and $(n+1)$-th bands is defined as
\begin{equation}
   {G_{n}} = \mathop {\min }\limits_{k_x,k_y}
  \left( {{E_{n + 1,{k_x},{k_y}}} - {E_{n,{k_x},{k_y}}}} \right).
\end{equation}
Due to the chiral symmetry, the two energy gaps for the purely off-diagonal AAH model are the same.
By solving a simple cube equation (see online supporting information), we analytically determine the transition point $\nu_{od}/J=4$, where the two energy gaps simultaneously close.
However, the chiral symmetry is broken and the transition point is split into two points even when a weak on-site potential modulation is applied (see online supporting information).
Accompanying with the topological phase transition, the gap in the energy  bands will close and reopen, see Fig.~\ref{gap} (b)-(d).
If the hopping modulation is sufficiently strong, $\nu_{od}/J>4$, a novel topological phase with large Chern numbers $(2,-4,2)$ appears.
Through the transition point, the Chern numbers change from $(-1,2,-1)$ to $(2,-4,2)$, where not only the values of Chern numbers become doubled but also their signs change.
This means that the particle pumping along the lattice direction becomes more fast and the propagation direction changes opposite.
In the region of $\nu_{od}/J > 4$, the energy gaps linearly increase with $\nu_{od}/J$, which may facilitate the detection of large Chern numbers in experiments.
The novel topological phases with large Chern numbers are not limited to the system of $\beta=1/3$, but generally exist in systems of rational $\beta$ and odd $q$ (see online supporting information).

\begin{figure}[!htp]
\begin{center}
\includegraphics[width=1\columnwidth]{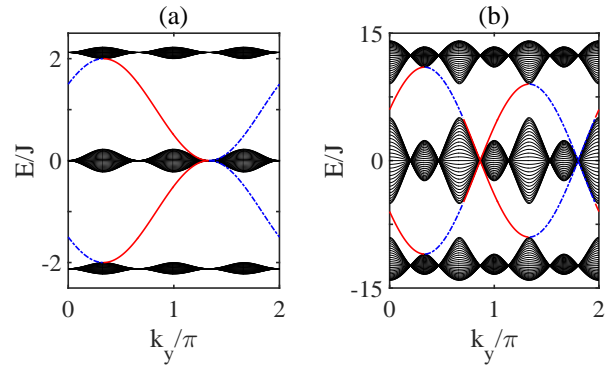}
\end{center}
\caption{ Energy spectrum for the off-diagonal AAH model with open boundary condition and different values of $\nu_{od}/J=$: (a) $1$ and (b) $10$. The blue dot-dashed (red solid) lines in the gaps stand for the edge modes localized at the left (right) edge. Here the lattice size is chosen as $89$ sites.}
\label{edge_state}
\end{figure}

Based upon the bulk-edge correspondence~\cite{Hatsugai1993a,Hatsugai1993b}, the topological features implied in the bulk state spectrum can be extracted in the topological edge state spectrum,
and the bulk-state Chern number equals the edge-state winding number difference between the neighboring gaps.
By implementing exact diagonalization, we obtain the energy spectrum for the off-diagonal AAH model with open boundary condition, see Fig.~\ref{edge_state} (a) for $\nu_{od}/J=1$ and (b) for $\nu_{od}/J=10$.
Due to the chiral symmetry, these energy spectra are mirror symmetric about zero energy.
For $\nu_{od}/J=1$, there are two edge modes with one at the left edge and the other at the right edge.
While for $\nu_{od}/J=10$, there are four edge modes with two at the left edge and the other two at the right edge.
This means that the corresponding winding numbers $(I_1, I_2)$ are $(-1,1)$ for $\nu_{od}/J=1$ and $(2,-2)$ for $\nu_{od}/J=10$.
As $I_0=0$ and $I_3=0$, the corresponding Chern numbers $(\mathcal C_1=I_1-I_0, \mathcal C_2=I_2-I_1, \mathcal C_3=I_3-I_2)$ respectively equal $(-1, 2, -1)$ and $(2, -4, 2)$, which are consistent with our numerical results shown in Fig.~\ref{phase_transition}.
Alternatively, the Chern numbers can also be deduced from the winding numbers of the reflection coefficient phase~\cite{Poshakinskiy2015,Mei2015}.

\section{Measuring Chern numbers via Thouless pumping of light}
\label{ssec:headsec}
We propose to measure the Chern numbers of the generalized AAH model via Thouless pumping of light~\cite{Thouless1983,Lohse2015,Nakajima2016}.
The abstract figure shows a vivid picture of Thouless pumping of light in the modulated waveguide arrays.
To implement Thouless pumping in photonic waveguide arrays, in addition to the periodic modulation of refractive indices along the lattice direction, one has also to modulate the refractive indices adiabatically and periodically along the propagation direction, shown in Fig.~\ref{diagram} (a).
Such waveguide arrays have been designed in fused silica by the femtosecond-laser written technique~\cite{Szameit2006,Szameit2009,Kraus2012a}.
The refractive index modulation can be realized by slightly varying the writing velocity along the propagation direction.

The light propagation is described by a paraxial filed $\psi (x,z)$ governed by the Schr\"{o}dinger equation with the longitudinal propagation distance $z$ acting the role of the evolution time,
\begin{equation}
  i\partial_z \psi(x,z)=-{1\over{2k_0}}{\partial^2\over{\partial x^2}}\psi(x,z)-{k_0\gamma R(x,z) \over{n_0}}\psi(x,z).
\end{equation}
Here, $\gamma$ describes the refractive index lattice depth and $k_0=2\pi n_0/\lambda$ (where we choose the refractive index $n_0=1.45$ and the wavelength $\lambda=0.63\mu m$).
The refractive index profile of the waveguide arrays is given by
\begin{equation}
  R(x,z)=\sum\limits_{j}\left[1+\alpha \cos(2\pi\beta j+\Omega z)\right]e^{-{(x-jw_s)^6/{w_x^6}}},
\end{equation}
where the super-Gaussian refractive index profile of individual waveguides is characterized by the normalized width $w_x=3\mu m$ and the waveguide spacing $w_s=10\mu m$.
The refractive index of $j$-th waveguide is modulated along the propagation direction with modulation strength $\alpha$, phase $2 \pi \beta j $, and frequency $\Omega=2\pi/Z$ (where $Z$ is the modulation period).
Obviously, such a refractive index lattice acts as a spatially periodic lattice potential with temporally periodic on-site modulations.
In our simulation, the parameters are chosen as $\alpha=0.5$, $\beta=p/q=1/3$.
The lattice depth are set as $\gamma=9\times 10^{-4}$ and $5\times 10^{-4}$ with the corresponding propagation distance $Z=30cm$ and $10cm$, respectively.
If the lattice depth $\gamma$ is sufficiently large and the modulation strength $\alpha <1$, one can expand the light field
\begin{equation}
\psi(x,z)=\sum_{j} C_{j} (z) \Phi_{j} (x,z)
\end{equation}
with lowest Wannier states $\Phi_{j} (x,z)$, where the expansion amplitudes $C_{j}(z)$ are complex numbers.
From Eq.~(5), by neglecting the constant on-site refractive index which only gives a uniform phase for $C_j (z)$, one can find that the evolution of $C_{j} (z)$ obeys the AAH model~\eqref{GAAH},
\begin{gather}
i\partial_z C_j (z)= - \left[J_{j,j+1} C_{j+1} (z)+ J_{j,j-1}C_{j-1} (z) \right] \nonumber\\
~~~~~~~~~~~~+ \nu_d\cos\left[{2\pi \over{3}}j+k_y(z)\right]C_{j}(z),
\end{gather}
with the parameters $(J, \nu_{od}, \nu_d)$ given by overlap integrals of Wannier states $\Phi_j(x,z)$, $\delta\phi=\pi/3$, and $k_y(z)=\Omega z$.
More details for the Wannier expansion are given in our online supporting information.

We have calculated the parameters for different values of the lattice depth $\gamma$.
Given $\gamma=9\times 10^{-4}$, we have the parameters $J=3.76\times 10^{-4}/\mu m$, $\nu_{od}=1.43\times 10^{-4}/\mu m$, $\nu_{d}=-36.06\times 10^{-4}/\mu m$.
Given $\gamma=5\times 10^{-4}$, we have the parameters $J=5.23\times 10^{-4}/\mu m$, $\nu_{od}=1.01\times 10^{-4}/\mu m$, $\nu_{d}=-18.34\times 10^{-4}/\mu m$.
Obviously, for these parameters, the off-diagonal modulation strengths $\nu_{od}$ are much smaller then the diagonal ones $\nu_d$.
To decrease the off-diagonal modulation strength $\nu_{od}$, one has to decrease the modulation strength $\alpha$ in further.
When the modulation strength $\alpha$ becomes not too strong, the off-diagonal modulation $\nu_{od}$ becomes far less than the hopping constant $J$.
Although the off-diagonal modulation is not completely switched off, our numerical results show that it does not change the efficiency of the Thouless pumping of light in the continuous model.

To realize adiabatic pumping, $k_y (z)$ has to be tuned slowly to suppress the Landau-Zener (LZ) transition between different bands.
According to the LZ formula~\cite{Landau1981}, the ratio of LZ transition from the first band to the second band is approximately given as $\Gamma\sim \exp\left(- G_1^2 Z\right)$.
One has to set $Z$ large enough to minimize $\Gamma$, but longer waveguide always causes larger total loss in realistic photonic systems.
The loss at the waveguide length $Z=10cm$ can be neglected in experiments while the loss at the waveguide length $Z=30cm$ may limit the observation of Thouless pumping of light~\cite{Szameit2009}.
To reduce the waveguide length, one can alternately design $\Omega$ varying with the energy band gap instead of linearly controlling $k_y(z)$.
To implement Thouless pumping, the initial states have to uniformly occupy the energy band in the quasi-momentum space.
However, photons can not automatically fill the energy band in a uniform way as fermions which follow the Pauli exclusion principle.
Generally, the initial state can be chosen as the Wannier state of the first band~\cite{Marzari1997,Marzari2012}, which can be realized by the fractional Fourier transformation~\cite{Leija2015}.
We find that a flat energy band can make the Wannier state more localized, and the maximally localized Wannier state of the first band can be approximately prepared by injecting a Gaussian laser beam in a single waveguide with largest refractive index, see the cover figure.
The input conditions are $A\exp(-x^2/W^2)$ with the width $W=3.77\mu m$ for the waveguide length $Z=30cm$ and $W=4.47\mu m$ for $Z=10cm$.

Based upon the continuous Schr\"{o}dinger equation~(5) for the light field $\psi(x,z)$, the parameter $k_y$ increases from $0$ to $2\pi$ with $z$ increasing from $0$ to $Z$ in one pumping cycle, the Chern number $\mathcal C$ can be derived from the mean transverse position shift of the light with the relation
\begin{equation}
 \mathcal C = {1\over{q w_s}}\left(\int {x|\psi(x,Z)|^2 dx }-\int {x|\psi(x,0)|^2 dx }\right).
  \label{TPC}
\end{equation}
We use spectral split method to simulate the light propagation in the modulated waveguide arrays~\cite{Feit1982}.

\begin{figure}[htp]
\begin{center}
\includegraphics[width=1\columnwidth]{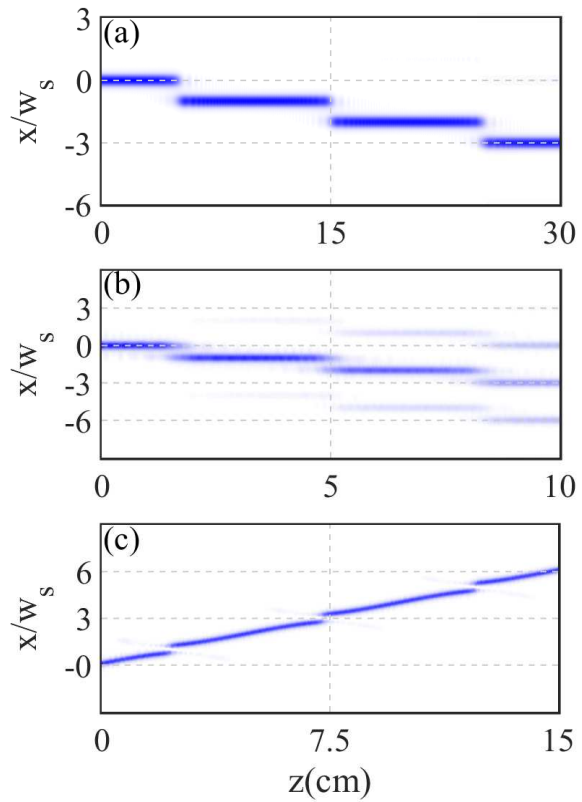}
\end{center}
\caption{The light intensity distributions $|\psi(x,z)|^2$ in one pumping cycle with different modulations:
(a) the refractive index modulation (6) with $\alpha=0.5$, $\beta=1/3$, $w_s=10\mu m$, $w_x=3\mu m$, $\gamma=9\times 10^{-4}$ and $Z=30cm$,
(b) the refractive index modulation (6) with $\alpha=0.5$, $\beta=1/3$, $w_s=10\mu m$, $w_x=3\mu m$, $\gamma=5\times 10^{-4}$ and $Z=10cm$,
and (c) the inter-waveguide spacing modulation (10) with $\beta=1/3$, $w_s=20\mu m$, $w_x=3\mu m$, $w_m=18\mu m$,  $\gamma=5\times10^{-4}$, $Z=15 cm$ and $ \phi_0=\pi/5$, .}
\label{pump}
\end{figure}

In Fig.~\ref{pump} (a) and (b), we show the light intensity distributions $|\psi(x,z)|^2$ for different values of the lattice depth $\gamma$ and the propagation distance $Z$.
The Chern numbers extracted from the mean position shifts are -0.97 for case (a) $(\gamma=9\times 10^{-4}, Z=30cm)$ and -0.99 for case (b) $(\gamma=5\times 10^{-4}, Z=10cm)$.
Both of them are almost as same as the ideal value -1 for the perfectly infinite system.
This means that the Thouless pumping of light provides another hallmark of the topological phases in our generalized 1D AAH photonic waveguide arrays.
Similarly, one can also determine the Chern numbers for excited bands by inputting maximally localized Wannier states occupying the corresponding bands.
In Fig.~\ref{pump} (a), the intensity distributions are mostly restricted in a single waveguide in the pumping process. This is because larger lattice depth makes the energy bands more flat and the flat band can efficiently suppress the diffusion along the lattice direction.
When the lattice depth becomes lower, the energy bands become less flat and induce different group velocities with different momentum $k_x$.
In Fig.~\ref{pump} (b), although the mean position shift obeys the Thouless pumping governed by the corresponding Chern number, the wavepacket spreads to both sides of the mean position and it becomes diffusive.

To observe a large Chern number, one can implement Thouless pumping by periodically modulating the inter-waveguide separations along the transverse direction, as shown in Fig.~\ref{diagram} (b).
The refractive index profile of the waveguide arrays is given by
\begin{equation}
  R(x,z)=\sum\limits_{j}e^{-{(x-x_j)^6/{w_x^6}}},
\end{equation}
where $x_j=jw_s+w_m \cos(2\pi \beta j+\Omega z+\phi_0)$ is the $j$-th waveguide center, $w_m$ is the modulation amplitude and $\phi_0$ is the initial phase.
In our simulation, the parameters are chosen as $\beta=1/3,~w_s=20\mu m,~w_x=3\mu m,~w_m=18\mu m,~\gamma=5\times10^{-4},~Z=15 cm$ and $\phi_0=\pi/5$.
The initial state is prepared by injecting the Gaussian laser beam with $W=4.3\mu m$ into the waveguide with the largest inter-waveguide separations.
In Fig.~\ref{pump} (c), we show the light intensity distributions $|\psi(x,z)|^2$ in one pumping cycle.
The Chern numbers extracted from the mean position shifts are $1.97$, which is very close to the ideal value $2$.

\section{Summary and discussion}

We have predicted topological phase transitions in the generalized AAH photonic waveguide arrays and have demonstrated a possibility to measure directly their Chern numbers via Thouless pumping of light. For weak hopping modulations, the off-diagonal commensurate AAH model has been shown to be topologically equivalent to its diagonal counterpart. Strengthening the hopping modulation, the energy bands undergo a gap closure with topological phase transitions, and novel topological phases with large Chern numbers appear when the hopping modulation is sufficiently strong. Taking the off-diagonal AAH model with $\beta=1/3$ as an example, we have analyzed the topological phase transitions at $\nu_{od}/J=\pm 4$. The photonic topological phase transitions can be verified by either topological edge or bulk invariants. To clarify the bulk-edge correspondence, in addition to the calculation of the bulk-state Chern numbers, we have analyzed the topological edge states and their winding numbers, and have suggested a simple scheme to measure directly the bulk Chern numbers by means of the Thouless pumping of light.

We know, in electronic systems, large Chern number means high Hall conductance.
Similarly, in photonic waveguide arrays, large Chern number means high coupling efficiency between waveguides.
The search of large Chern numbers is a significant goal in exploring novel topological phases~\cite{Skirlo2014,Skirlo2015}.
Generally speaking, large Chern numbers are introduced by weak tunneling in high-order perturbation expansion~\cite{Andrei2013}.
Moreover, because of a narrow energy gap, large Chern numbers are very hard to measure in experiments. We believe that our findings of large Chern numbers in photonic waveguide arrays and the suggestion of their measurement via Thouless pumping offer an inspiration for experimental studies of novel topological phases.

Beyond linear photonic waveguide arrays, it is important to understand the interplay between nonlinearity and topology in nonlinear photonic waveguide arrays.
It is well-known that, in an interacting many-body quantum system, fractional topological states can be induced by the interplay between interaction and topology~\cite{Niu2010}.
In the mean-field theory, the inter-particle interaction can be viewed as a kind of nonlinear mean-field interaction and the system obeys a nonlinear Schr\"{o}dinger equation.
In the other hand, due to the nonlinearity, gap solitons may emerge in nonlinear quantum lattices~\cite{Chen1987,Kivshar2007,Torner2010}.
Therefore, it is deserved to study the exotic Thouless pumping of light (in particular the gap solitons) in nonlinear photonic waveguide arrays.

\begin{acknowledgement}
We thank J. Huang, D. Neshev, A. Poddubny, K. Wang, and L.~Zhang for useful discussions and suggestions. This work was supported by the National Basic Research Program of China (NBRPC) under Grant No. 2012CB821305, the National Natural Science Foundation of China (NNSFC) under Grants No. 11374375 and 11574405, and the Australian Research Council.
\end{acknowledgement}


\end{document}